\documentclass[preprint2]{aastex}	

\usepackage{natbib}
\usepackage{amssymb}
\usepackage[displaymath]{lineno}
\usepackage{graphicx}
\usepackage{subfigure}
\usepackage{lineno}

\usepackage{captcont}
\usepackage[final,		
	colorlinks,			
	linktocpage,		
	linkcolor=blue,		
	citecolor=blue,		
	urlcolor=blue,		
	breaklinks=true,	
	]{hyperref}
\usepackage[all]{hypcap} 

\shorttitle{Discovery of a blazar in the Galactic plane}
\shortauthors{[J.~Vandenbroucke et al.]}

\begin{document}


\title{Discovery of a GeV blazar shining through the Galactic plane}

\author{
J.~Vandenbroucke\altaffilmark{*,1}, 
R.~Buehler\altaffilmark{*,1}, 
M.~Ajello\altaffilmark{1}, 
K.~Bechtol\altaffilmark{1}, 
A.~Bellini\altaffilmark{2,3}, 
M.~Bolte\altaffilmark{4}, 
C.~C.~Cheung\altaffilmark{5,6}, 
F.~Civano\altaffilmark{7}, 
D.~Donato\altaffilmark{8}, 
L.~Fuhrmann\altaffilmark{9}, 
S.~Funk\altaffilmark{1}, 
S.~E.~Healey\altaffilmark{1}, 
A.~B.~Hill\altaffilmark{10,11}, 
C.~Knigge\altaffilmark{12}, 
G.~M.~Madejski\altaffilmark{1}, 
R.~W.~Romani\altaffilmark{1}, 
M.~Santander-Garc\'ia\altaffilmark{13,14,15}, 
M.~S.~Shaw\altaffilmark{1}, 
D.~Steeghs\altaffilmark{16}, 
M.~A.~P.~Torres\altaffilmark{7}, 
A.~Van~Etten\altaffilmark{1}, 
K.~A.~Williams\altaffilmark{17}
}
\altaffiltext{*}{Corresponding authors: J.~Vandenbroucke, justinv@stanford.edu; R.~Buehler, buehler@stanford.edu}
\altaffiltext{1}{W. W. Hansen Experimental Physics Laboratory, Kavli Institute for Particle Astrophysics and Cosmology, Department of Physics and SLAC National Accelerator Laboratory, Stanford University, Stanford, CA 94305, USA}
\altaffiltext{2}{Dipartimento di Astronomia, Universit\`a di Padova, I-35122 Padova , Italy}
\altaffiltext{3}{Space Telescope Science Institute, Baltimore, MD 21218, USA}
\altaffiltext{4}{Santa Cruz Institute for Particle Physics, Department of Physics and Department of Astronomy and Astrophysics, University of California at Santa Cruz, Santa Cruz, CA 95064, USA}
\altaffiltext{5}{Space Science Division, Naval Research Laboratory, Washington, DC 20375, USA}
\altaffiltext{6}{National Research Council Research Associate, National Academy of Sciences, Washington, DC 20001, USA}
\altaffiltext{7}{Harvard-Smithsonian Center for Astrophysics, Cambridge, MA 02138, USA}
\altaffiltext{8}{NASA Goddard Space Flight Center, Greenbelt, MD 20771, USA}
\altaffiltext{9}{Max-Planck-Institut f\"ur Radioastronomie, Auf dem H\"ugel 69, 53121 Bonn, Germany}
\altaffiltext{10}{Universit\'e Joseph Fourier - Grenoble 1 / CNRS, laboratoire d'Astrophysique de Grenoble (LAOG) UMR 5571, BP 53, 38041 Grenoble Cedex 09, France}
\altaffiltext{11}{Funded by contract ERC-StG-200911 from the European Community}
\altaffiltext{12}{School of Physics and Astronomy, University of Southampton, Highfield, Southampton, SO17 1BJ, UK}
\altaffiltext{13}{Instituto de Astrof\'isica de Canarias, E38205 - La Laguna (Tenerife), Spain}
\altaffiltext{14}{Isaac Newton Group of Telescopes, E-38700 Sta. Cruz de la Palma, Spain}
\altaffiltext{15}{Departamento de Astrofisica, Universidad de La Laguna, E-38205 La Laguna, Tenerife, Spain}
\altaffiltext{16}{Department of Physics, University of Warwick, Coventry CV4 7AL, UK}
\altaffiltext{17}{Department of Astronomy, University of Texas, Austin, TX 75712, USA}

\begin{abstract}
The \emph{Fermi} Large Area Telescope (LAT) discovered a new gamma-ray source near the Galactic plane, \object{Fermi J0109+6134}, when it flared brightly in 2010 February.  The low Galactic latitude ($b =$~-1.2\degr) indicated that the source could be located within the Galaxy, which motivated rapid multi-wavelength follow-up including radio, optical, and X-ray observations.  We report the results of analyzing all 19 months of LAT data for the source, and of X-ray observations with both \emph{Swift} and the \emph{Chandra X-ray Observatory}.  We determined the source redshift, $z =$~0.783, using a Keck LRIS observation.  Finally, we compiled a broadband spectral energy distribution (SED) from both historical and new observations contemporaneous with the 2010 February flare.  The redshift, SED, optical line width, X-ray absorption, and multi-band variability indicate that this new GeV source is a blazar seen through the Galactic plane.  Because several of the optical emission lines have equivalent width $>5$~\AA, this blazar belongs in the flat-spectrum radio quasar category.
\end{abstract}



(Dated: \today)



\section{Introduction}

Variability in gamma-rays, as in other wavebands, is an essential diagnostic for identifying sources and determining emission mechanisms.  Gamma-ray variability has been established for numerous extragalactic sources, all active galactic nuclei (AGN) and gamma-ray bursts.  Within the Galaxy, pulsars exhibit periodic variability on short (milliseconds to seconds) time scales corresponding to their spin period (and are otherwise relatively steady), and binary systems exhibit periodic variability on long (hours to days) time scales corresponding to their orbital period~\citep{1FGL}.

Accreting binary systems have exhibited transient outbursts during which their X-ray emission increases by a factor up to $\sim$10$^6$.  One particular X-ray binary, Cygnus X-3 (a microquasar), has exhibited flaring activity in GeV gamma rays that is correlated with radio outbursts~\citep{CygX3}.  Transient emission from Cygnus X-1 has also been claimed by \emph{AGILE}~\citep{CygX1}.

Flaring activity has not been established for Galactic GeV source classes other than binary systems.  However, the Energetic Gamma Ray Experiment Telescope (EGRET) on board the \emph{Compton Gamma Ray Observatory} did detect a bright gamma-ray transient near the Galactic plane, which had no radio-loud flat-spectrum radio counterpart and had properties inconsistent with those of an AGN or isolated pulsar~\citep{Tavani97}.  This source was never identified and has been interpreted as evidence for a new class of variable gamma-ray sources.

The unexplained EGRET event has motivated an ongoing search for new gamma-ray transients near the Galactic plane, as well as for variability among established Galactic gamma-ray sources.  Prior to the discovery reported here, three transients near the Galactic plane \citep{Hays09,atel2081} have been detected with the Large Area Telescope (LAT) on board the \emph{Fermi Gamma-ray Space Telscope}.  None of them has yet been identified.

On 2010 February 1\footnote{Dates and times are UTC throughout.}, the LAT~\citep{Atwood09} discovered a new, flaring source (\object{Fermi J0109+6134}) near the Galactic plane~\citep{atel2414}.  The \emph{AGILE} gamma-ray telescope confirmed the \emph{Fermi} LAT detection of the flare~\citep{atel2416}.  We report an analysis of the full 19-month LAT data set for the source region.  We also report new multi-wavelength observations, including an optical spectrum from which we have determined the source redshift.  Finally, we present a compilation of the broadband spectral energy distribution (SED) from both contemporaneous and historical data.  The redshift, optical line width, SED shape, and multi-band variability indicate a blazar identity for the source.

\begin{figure}[t!]
\centering
\noindent\includegraphics[width=0.48\textwidth]{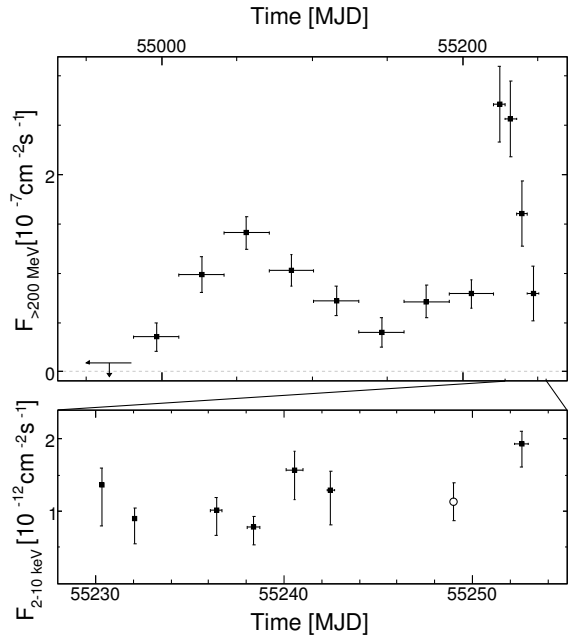}
\caption{Upper panel: \emph{Fermi} LAT light curve, for photons with energy above 200 MeV.  The source was not detected in a combined data set containing all events between 2008 August 4 (MJD 54682) and 2009 May 30 (MJD 54971).  This $\sim$10~month time interval is represented with a single upper limit.  Lower panel: 2--10 keV X-ray light curve.  There are seven \emph{Swift} XRT observations (filled squares) and one \emph{Chandra} observation (open circle).  X-ray fluxes are shown as measured, without accounting for absorption.  Vertical error bars are $\pm$1~$\sigma$ uncertainty (statistical only).}
\label{light_curve}
\end{figure}

\section{\emph{Fermi} LAT analysis}

We analyzed LAT data in a 15\degr~radius around \object{Fermi J0109+6134} for the full data set acquired to date (2008 August 4 through 2010 February 23).  We performed an unbinned likelihood fit to the data using the {\tt gtlike} tool, which is part of the \emph{Fermi} Science Tools\footnote{Version v9r15p4}.  The statistical significance of the background-plus-source model relative to the background-only model is quantified with the ``test statistic,'' TS = 2$\Delta$log(likelihood).  The unbinned likelihood method is described in~\citet{Mattox96} and~\citet{1FGL}.  The background model we used includes two diffuse components, one Galactic (model {\tt gll\_iem\_v02}) and one isotropic (model {\tt isotropic\_iem\_v02}), as well as nearby sources in the first \emph{Fermi} LAT catalog~\citep[1FGL;][]{1FGL}.

We fit \object{Fermi J0109+6134} with a power law energy spectrum, $\Phi(E) = \Phi_0 E^{-\Gamma_\gamma}$, with both the normalization $\Phi_0$ and gamma-ray photon index $\Gamma_\gamma$ free.  The normalization of each of the following background components was also free in the fit: each point source in 1FGL within a 6\degr~radius of \object{Fermi J0109+6134}, the nearby bright and variable binary LS~I~+61\degr303, the Galactic diffuse model, and the isotropic diffuse model.  Only those events belonging to the highest quality (``diffuse'') class and with energies between 200 MeV and 300 GeV were included.  Events with zenith angle greater than 105\degr~were removed to avoid contamination by gamma rays produced from cosmic ray interactions in the atmosphere~\citep{FermiAlbedo}.  We used the {\tt P6\_V3\_DIFFUSE} instrument response functions.

No statistically significant flux was detected above background at the location of \object{Fermi J0109+6134} in the 10-month LAT data set from 2008 August 4 through 2009 May 30:  The TS for the source hypothesis using all data in this time interval is  3.  The 95\% confidence upper limit on the flux of photons above 200~MeV is $9.3\times10^{-9}$ photons cm$^{-2}$s$^{-1}$.

For the following 9-month interval, from 2009 May 30 through 2010 February 23, \object{Fermi J0109+6134} was detected with a TS of 639, corresponding to a significance of $\sim$25~$\sigma$.  The integral flux in this interval is $(0.93\pm0.05)\times10^{-7}$ photons cm$^{-2}$s$^{-1}$ above 200~MeV, with $\Gamma_\gamma=2.59\pm0.06$.  Uncertainties are statistical only.  According to the best-fit source model, 1577 photons with energy greater than 200~MeV are attributed to this source in the 9-month time interval.  From this analysis, the best-fit LAT position\footnote{We use the J2000.0 epoch throughout.} is (\emph{RA}~=~01$^{\rm h}$09$^{\rm m}$58\fs4, \emph{Dec}~=~$+$61\degr32\arcmin58\arcsec).  The LAT 68\% error radius is 1.8\arcmin, corresponding to a 95\% error radius of 3.0\arcmin.

No GeV source was previously known at this location.  It was not detected in the EGR~\citep{EGR}, 3EG~\citep{3EG}, or 1FGL~\citep{1FGL} catalog.  The nearest source in 1FGL is 1FGL J0131.2+6121, an unassociated source 2.55\degr~away~\citep{1FGL}.

The data from the 9-month interval were divided into monthly time bins, and a likelihood analysis was performed in each of them.  The source was detected significantly (TS$>$10) in each month.  The gamma-ray light curve is shown in~Figure~\ref{light_curve}.  The source is clearly variable on a $\sim$1 month time scale, with two distinct flares peaking in 2009 August and 2010 February.  The flux was at least 30 times greater at the peak of the flare than when it was not detected prior to the flare.  The power law index was determined for each time bin and showed no significant evidence of variation in the 9-month analysis.

\begin{figure*}[t]
\begin{center}
\subfigure[][]{
\label{zoom_out}
\noindent\includegraphics[width=0.57\textwidth]{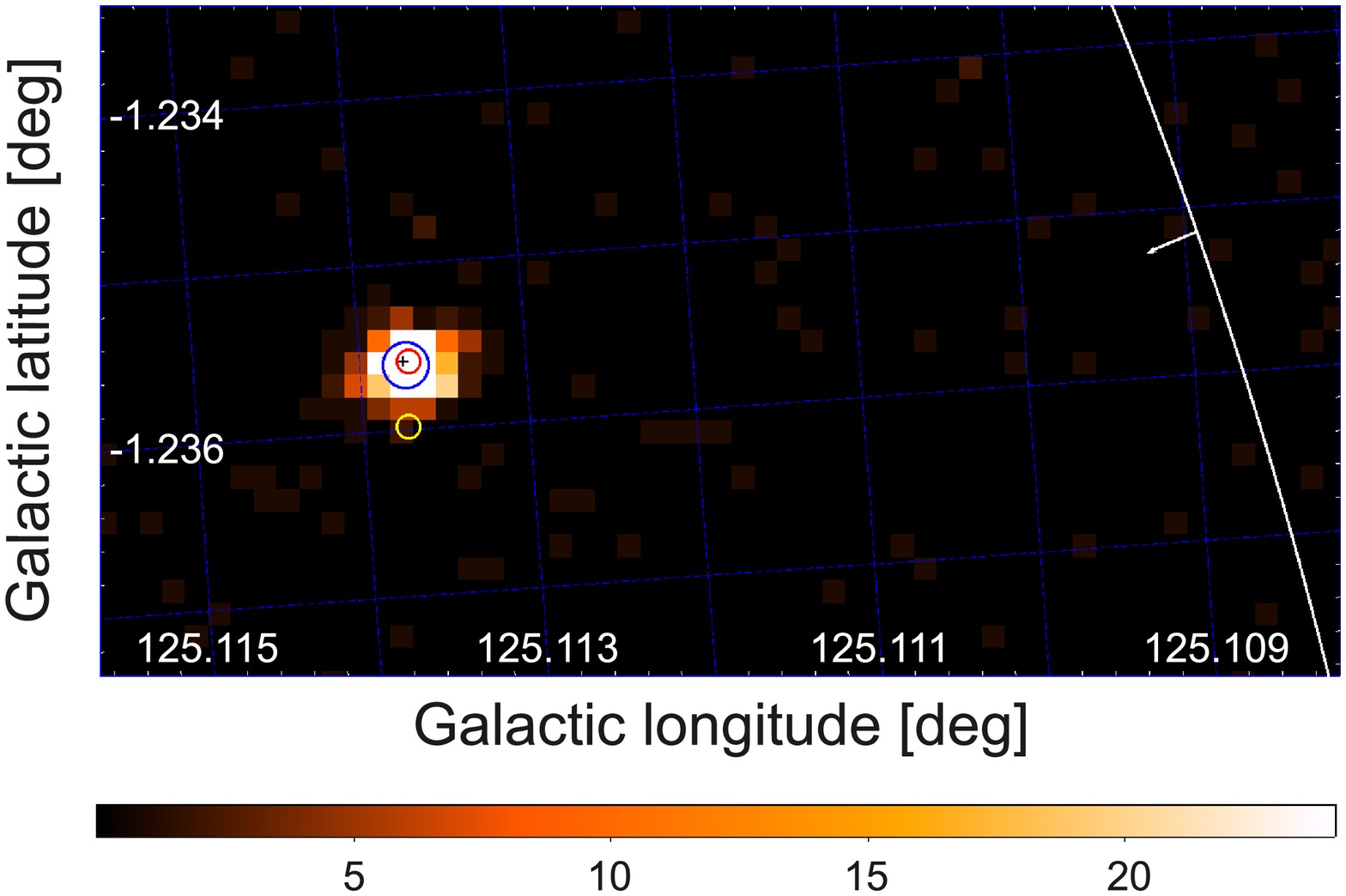}
}
\subfigure[][]{
\label{zoom_in}
\noindent\includegraphics[width=0.38\textwidth]{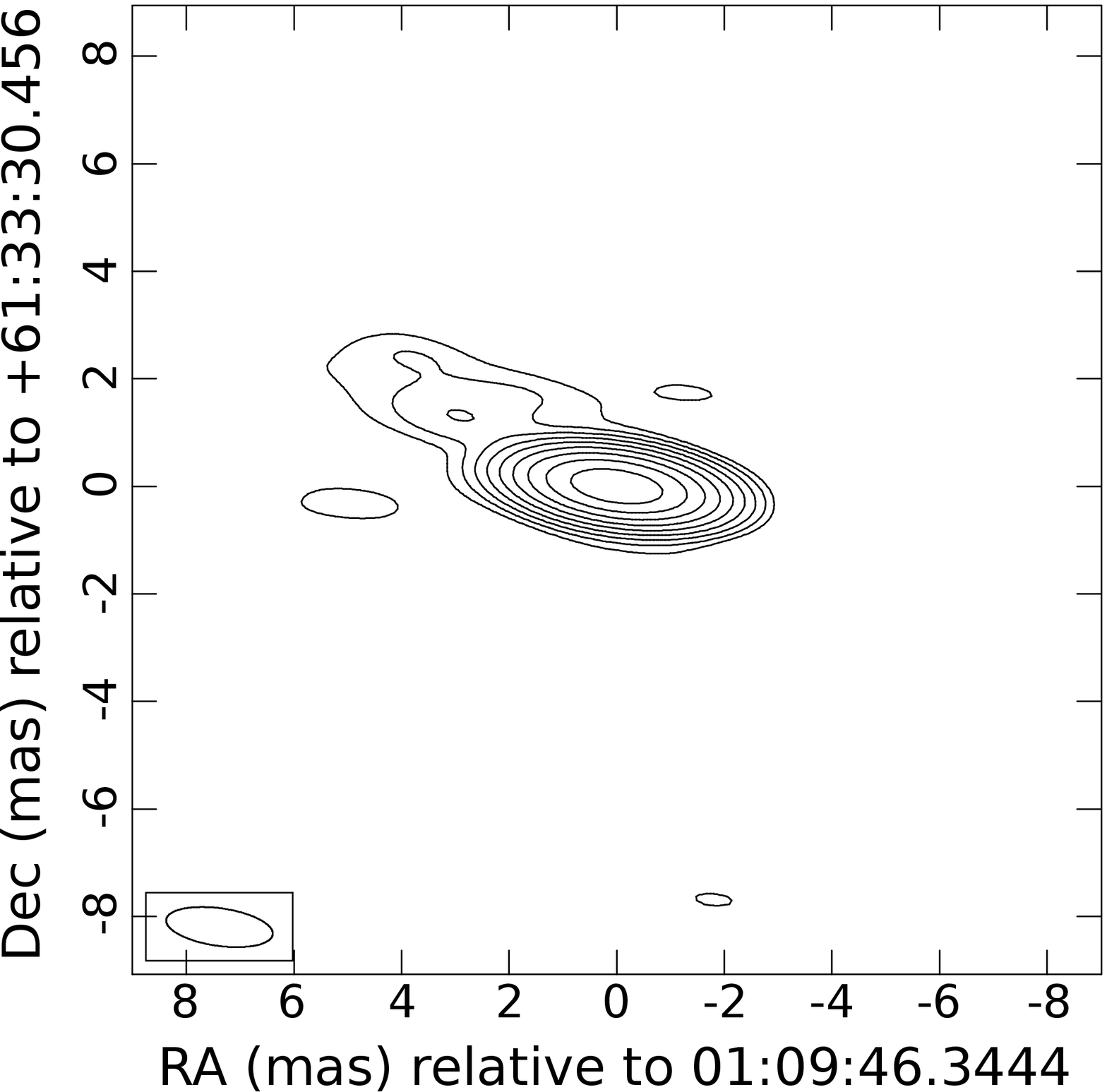}
}
\caption{\subref{zoom_out} Gamma, X-ray, optical, and radio counterpart positions.  The background image is a counts map from the \emph{Chandra} observation.  The \emph{Fermi} LAT 68\% error circle is indicated by the white arc. The \emph{Fermi} LAT best fit position is outside the figure, at ($l=125.138\degr, b=-1.243\degr$), in the direction of the white arrow.  The two optical candidate counterparts are indicated by yellow (Paredes et al. counterpart) and red (IPHAS counterpart) circles.  The two optical error radii are estimated at 0\farcs25.  The VCS2 radio position is indicated with a black `+,' and the best-fit \emph{Chandra} position is indicated with a blue circle of radius 0\farcs5.  Both the optical and \emph{Chandra} position uncertainties are dominated by systematic frame offsets.  The radio and X-ray positions confirm that the IPHAS source, rather than the Paredes et al. source, is the correct optical counterpart.  \subref{zoom_in} Reprocessed VLBA 8.6 GHz image from VCS2~\citep{Fomalont03} showing a jet extension of order 10~pc. The beam (2 mas~$\times$~0.7 mas at position angle = 82\degr) is indicated to the lower left and the contour levels begin at 1.5 mJy/bm and increase by factors of 2.}
\label{counterparts}
\end{center}
\end{figure*}

\section{Counterparts}

We have identified radio, optical, and X-ray counterparts of the new GeV gamma-ray source.  Their positions are shown in Figure~\ref{counterparts}\subref{zoom_out}.

\subsection{Radio}

In archival radio data we identified a likely flat-spectrum radio counterpart to the new GeV source: \object{GT 0106+613} (\object{VCS2 J0109+6133}).  This is the only radio source known within the 68\% LAT error radius.  This source was first discovered by Gregory and Taylor in their radio survey of the northern Galactic plane~\citep{Taylor83, Gregory86}, performed with the NRAO 91-m telescope at 5~GHz.  They determined that it had compact morphology, in comparison with other sources that featured doublet or triplet morphology indicating one or two radio lobes in addition to a central object.  A precise position for the radio source was later determined as part of the VCS2 very long baseline interferometry survey~\citep{Fomalont03}: (\emph{RA}~=~01$^{\rm h}$09$^{\rm m}$46\fs34439~$\pm$~0\fs00013, \emph{Dec}~=~$+$61\degr33\arcmin30\farcs4557~$\pm$~0\farcs0003).

The high-resolution VCS2 image resolved few-milli-arcsec (few-parsec, see redshift determination below) jet extension~\citep{Fomalont03}.  This image\footnote{\url{http://astrogeo.org/vcs2/vcs2_cat.html}} is shown in Figure~\ref{counterparts}\subref{zoom_in}.

\subsection{Optical}

\citet{Paredes93} followed up the variable radio sources detected by Gregory and Taylor with optical observations.  They detected a candidate optical counterpart of \object{GT 0106+613} with an \emph{I} band magnitude of 19.4 and position (\emph{RA}~=~01$^{\rm h}$09$^{\rm m}$46\fs33, \emph{Dec}~=~$+$61\degr33\arcmin29\farcs1), 1.4\arcsec~offset from the VCS2 radio position.  However, without a spectrum they were unable to determine whether the source was Galactic or extragalactic.

In addition to the Paredes et al. observations performed in 1992, the source lies within both the IPHAS~\citep{Drew05}\footnote{\url{http://www.iphas.org}} and UVEX~\citep{Groot09} optical surveys of the northern Galactic plane.  The IPHAS images (obtained between 2003 and 2006) show the Paredes et al. source, and they also show another optical source (\object{IPHAS J010946.33+613330.5}) which matches the radio position better than the Paredes et al. candidate.  This source has a position (\emph{RA}~=~01$^{\rm h}$09$^{\rm m}$46\fs33, \emph{Dec}~=~$+$61\degr33\arcmin30\farcs5), 0\farcs1~from the VCS2 radio position.  The IPHAS and UVEX magnitudes are ($i' = 19.7, r'-i' = 1.2$) for the Paredes et al. counterpart, and ($i' = 19.7, r'-i' = 1.5$) for the IPHAS counterpart that is closer to the radio position~\citep{atel2429}.

\subsection{X-ray}

Following the gamma-ray flare, we obtained a series of seven observations of the source region with the \emph{Swift} X-ray Telescope (XRT)\footnote{\emph{Swift} Target ID 31604.} between 2010 February 3 and 2010 February 25, each with a duration between 5 and 10~ksec.  We obtained a 10~ksec \emph{Chandra} ACIS-S3 (AXAF CCD Imaging Spectrometer) observation, performed\footnote{\emph{Chandra} Observation ID 11685, Sequence Number 11685.} on 2010 February 21-22.

In the initial \emph{Swift} XRT observation~\citep{atel2420}, two X-ray sources were detected in the full XRT field.  One was outside the LAT 95$\%$ error circle.  The other was a bright source positionally coincident with the radio candidate, with (\emph{RA}~=~01$^{\rm h}$09$^{\rm m}$46\fs86, \emph{Dec}~=~$+$61\degr33\arcmin29\farcs3) and a 90$\%$ error radius of 4\farcs4 according to the initial 5~ksec observation.

Figure~\ref{counterparts}\subref{zoom_out} shows the X-ray counts map obtained with \emph{Chandra}.  In the \emph{Chandra} field of view (larger than the zoom shown here), there are several sources detected.  One matches the VCS2 radio position well.  The next closest is significantly dimmer and is 3.9\arcmin~away from the radio position, outside the LAT 95\% error circle.  The best-fit position of the \emph{Chandra} source corresponding to the radio position is (\emph{RA}~=~01$^{\rm h}$09$^{\rm m}$46\fs338, \emph{Dec}~=~$+$61\degr33\arcmin30\farcs42), with negligible statistical uncertainty that is dominated by a frame offset uncertainty of $\sim$0\farcs5.  There is no evidence for extension or jet morphology in the \emph{Chandra} image.


The Galactic column density in this direction is estimated to be $N_{\rm H} = 5.5\times10^{21}$~cm$^{-2}$ from to the Leiden/Argentine/Bonn (LAB) 21~cm survey~\citep{Kalberla05}.  We fit the \emph{Chandra} X-ray spectrum with a hard ($\Gamma_X=1.19^{+0.16}_{-0.15}$) absorbed ($N_{\rm H}=9.2^{+1.5}_{-1.3}\times10^{21}$~cm$^{-2}$) power law, with $\chi^2 = 40.3$ for 61-3=58 degrees of freedom.  It is not surprising that the column density estimated from the X-ray absorption is significantly larger than that estimated from the LAB radio survey data, particularly for this direction within 10\degr~of the Galactic plane.  The X-ray estimate is likely more reliable than the LAB estimate due to uncertainty in the neutral hydrogen spin temperature, which is used to infer column density from the 21~cm data.

The measured (absorbed) \emph{Chandra} flux is $1.15\times10^{-12}$~erg~cm$^{-2}$s$^{-1}$ in the 2--10~keV band.  The model (unabsorbed) flux is $1.23\times10^{-12}$~erg~cm$^{-2}$s$^{-1}$ in the same band.  The absolute flux uncertainty is $\sim25\%$.  We also fit the spectrum including a thermal component (a kT=1.5~keV MEKAL spectrum) fixed at the newly determined (see below) redshift, z=0.783.  Although some excess counts appear near the expected Fe K$\alpha$ line and near other lines corresponding to elements of intermediate mass, this thermal component was only detected at 1.4~$\sigma$ significance.

We analyzed archival \emph{Swift} Burst Alert Telescope (BAT) data from 2004--2009 and found that the source was detected with a significance of $\sim$4.5~$\sigma$.  We used this data to estimate the spectrum in the 14--195~keV band, following the procedure described in~\citet{Ajello08, Ajello09b}.  The resulting spectrum is accurate to the mCrab level~\citep{Ajello09a}.





\begin{figure}[t!]
\centering
\noindent\includegraphics[width=0.48\textwidth]{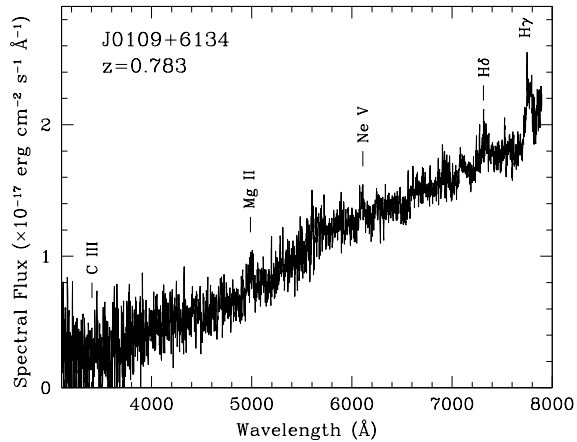}
\caption{Optical spectrum obtained with Keck.  Note that extinction has not been corrected.}
\label{keck_spectrum}
\end{figure}

\section{Optical spectrum and redshift}

We obtained an optical spectrum on 2010 February 9 using the Low Resolution Imaging Spectrometer (LRIS) with the atmospheric dispersion corrector~\citep{Oke95} at the Keck 1 telescope.  A 1\arcsec-long slit was placed at a position angle of 12.5 degrees east of north in order to include both optical sources in a single pointing.  The observations were performed at high airmass and as a consequence the image quality full-width half maximum (FWHM) was $\sim$1.2\arcsec.  The blue arm of the spectrograph used a 400 l mm$^{-1}$ grism blazed at 4000~\AA ~for a resolution of $\sim$7~\AA~FWHM; the upgraded red side of the spectrograph used a 600 l mm$^{-1}$ grating blazed at 7500~\AA, resulting in a spectral resolution of $\sim$4.7~\AA~FWHM.

Data reduction was performed with the IRAF package using standard techniques: an optimal extraction algorithm to maximize the signal-to-noise ratio (SNR); wavelength calibration from sky lines; spectrophotometric calibration and telluric subtraction from observations of G191-B2B; and visual inspection to clean cosmic rays. Multiple exposures were combined into a single spectrum, weighted by the SNR.

The Paredes et al. candidate showed Balmer absorption at rest, indicating that it is a field star.  The IPHAS counterpart showed a red continuum-dominated spectrum, with broad UV emission lines consistent with a redshifted blazar subject to Galactic extinction.

The spectrum of the latter source is shown in Figure~\ref{keck_spectrum}.  Five emission lines were identified: C~III, Mg~II, Ne~V, H$\delta$, and H$\gamma$.  From the spectrum we determine the redshift of this source to be 0.783.  The equivalent width of each of the Mg~II, Ne~V, H$\delta$, and H$\gamma$ lines exceeds 5~\AA.  The signal-to-noise ratio of the C~III line is too low to estimate its width.  Because the strongest emission line width was greater than 5~\AA, we classify this blazar as belonging to the flat-spectrum radio quasar (FSRQ) category rather than the BL Lacertae (BL Lac) category, following the scheme of the Candidate Gamma-Ray Blazar Survey~\citep{CGRaBS} and the first \emph{Fermi} LAT AGN catalog~\citep{1LAC}.

A fit to the continuum in the Keck spectrum (between 380~nm and 800~nm) gives $\nu F_{\nu}=8.3^{+3.6}_{-3.2}\times10^{-13}$~erg cm$^{-2}$ s$^{-1}$ at $5.45\times10^{14}$~Hz (550~nm) and $\alpha-1 = 0.1^{+0.3}_{-0.4}$, where $\nu F_{\nu} \propto \nu^{-(\alpha-1)} $ and the optical photon index is $\Gamma_{opt} = \alpha+1$.  Galactic extinction has been removed using the maps of~\citet{Schlegel98}.  The uncertainty on the normalization and index include 10$\%$ uncertainty in the extinction maps and 30$\%$ uncertainty in the absolute spectrophotometry due to slit loss and nearby source confusion.

\begin{figure*}[t!]
\centering
\noindent\includegraphics[width=1\textwidth]{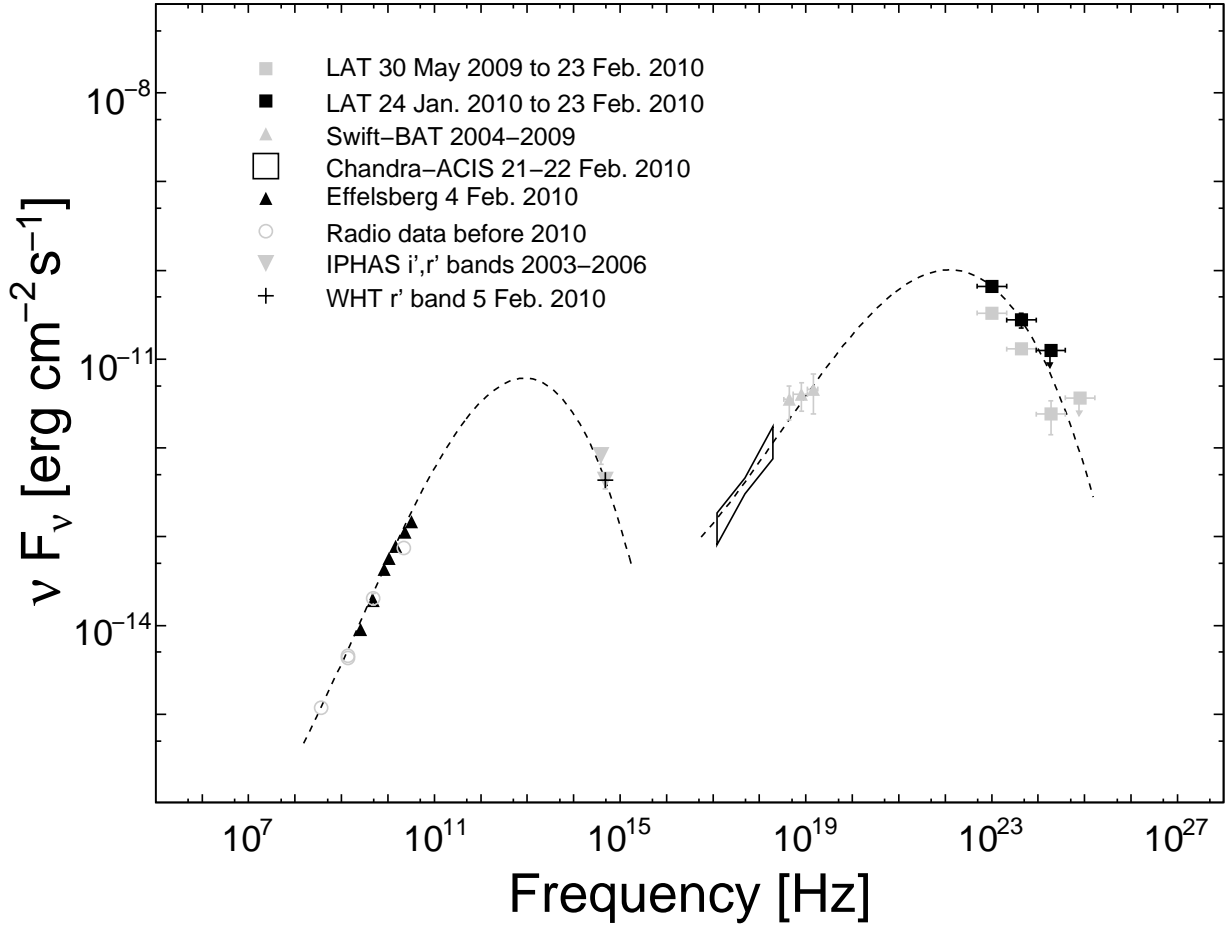}
\caption{Spectral energy distribution.  Historical data are shown in grey.  Simultaneous data acquired during the February 2010 flare are shown in black.  Extinction has been removed from both the optical and the X-ray spectra.  The \emph{Fermi} LAT spectrum was determined with a likelihood analysis analogous to the one described in the text, performed independently in each energy bin.  For energy bins with no statistically significant flux detected ($TS<9$), a 95\% upper limit was calculated and is indicated with arrows.  Historical radio data are from \citet{Douglas96}, \citet{White92}, \citet{Condon98}, \citet{Becker91}, and \citet{Petrov07}.  Dashed lines indicate a third-order polynomial fit to each bump.}
\label{sed}
\end{figure*}

\section{Radio, optical, and X-ray variability}

\citet{Taylor83} determined that this radio source exhibited long-term ($\geq1$~yr) and possibly short-term ($<1$~yr) variability.  Only a few percent of the sources in their northern Galactic plane survey exhibited variability, and they requested additional observations to determine which if any were Galactic.  \citet{Duric88} established that the source exhibited short-term as well as long-term radio variability.

In addition to historical (pre-2010) radio flux densities measured at several frequencies between 0.365 and 22~GHz, a new radio spectrum, contemporaneous with the 2010 February flare, was measured with the Effelsberg 100-m telescope on 2010 February 4.  The flux density was measured at several frequencies between 2.6 and 32~GHz.  At 8 (22)~GHz, the flux density was greater than archival values by $\sim$20\% ($\sim$50\%)~\citep{atel2428}, indicating radio flaring behavior contemporaneous with the gamma-ray flare.

There is marginal evidence for optical variability over the several years of IPHAS frames.  The absence of the IPHAS source in the Paredes et al. observations (when the Paredes et al. counterpart was detected clearly) provides stronger evidence of optical variability.  Both candidate counterparts were imaged using a 120~s exposure of the ACAM camera on the William Herschel Telescope (WHT) on 2010 February 5 to check the brightness at the time of the observed gamma-ray flare.  The Paredes et al. candidate had $r' = 20.9$ and the IPHAS candidate had $r' = 21.2$, each the same as determined from the 2003-2006 IPHAS images~\citep{atel2429}.

Figure~\ref{light_curve} shows the X-ray light curve compiled from the \emph{Swift} and \emph{Chandra} observations.  In the 2010 February 3--25 time interval, the X-ray flux shows no significant variability.


\section{Spectral energy distribution and interpretation}

We combined both historical and contemporaneous radio, optical, X-ray, and gamma spectra in the SED shown in Figure~\ref{sed}. The two-bump shape is consistent with typical blazar SED's where the low-energy bump is interpreted as synchrotron emission and the high-energy bump is interpreted as inverse Compton emission.

For reference, we fit a parameterization (third degree polynomial) to each bump, following the method of~\citet{FermiBlazarSEDs}.  This fit gives a synchrotron peak at $\sim$10$^{13}$~Hz and a Compton peak at $\sim$10$^{22}$~Hz, making this a Low Synchrotron Peaked (LSP) blazar following the classification scheme of ~\citet{FermiBlazarSEDs}.


The redshift, SED shape, optical line width, and multi-band variability of this source indicate that it is a background blazar seen shining through the Galactic plane. It provides an example of extragalactic GeV source identification and characterization in the difficult region very close to the Galactic plane.  Blazar catalogs generally avoid the region within $\pm$10\degr of the Galactic plane due to the significant extinction and source confusion in this region.

X-ray measurements of background blazars can improve our understanding of the Galactic interstellar medium.  With improved statistics, the absorption fit to the X-ray spectrum may constrain both the column density and composition in the direction of this source, independent of the LAB 21~cm survey.  Blazars are believed to have negligible intrinsic absorption~\citep{Perlman05}.

This source provides the first example of successfully determining the identity of a transient GeV source discovered near the Galactic plane by following up with collaborative multi-observatory, multi-wavelength observations.  The counterparts have been unambiguously identified thanks to the precise optical and X-ray positions.  Monitoring of this blazar is ongoing with the \emph{Fermi} LAT and other facilities.  The three other Galactic plane transients detected with the \emph{Fermi} LAT in its first 19 months of operation remain unidentified.  Observation and analysis are ongoing to identify these transients and to detect new GeV transients near the Galactic plane.  With its large field of view and frequent full-sky scans, Fermi is very well suited to discover more GeV transients across the sky and in particular in the Galactic plane.  Fermi observations of these transients will likely lead to new understanding of known GeV source classes as well as the discovery of new classes of GeV emitters.

\acknowledgments

We would like to thank the \emph{Chandra} and \emph{Swift} teams for timely and high quality X-ray observations.  We are grateful for valuable comments on the manuscript from Berrie Giebels.  J.~V. is supported by a Kavli Fellowship from the Kavli Foundation.  The $Fermi$ LAT Collaboration acknowledges support from a number of agencies and institutes for both development and the operation of the LAT as well as scientific data analysis. These include NASA and DOE in the United States, CEA/Irfu and IN2P3/CNRS in France, ASI and INFN in Italy, MEXT, KEK, and JAXA in Japan, and the K.~A.~Wallenberg Foundation, the Swedish Research Council and the National Space Board in Sweden. Additional support from INAF in Italy and CNES in France for science analysis during the operations phase is also gratefully acknowledged.



{\it Facilities:} \facility{Fermi (LAT)}, \facility{CXO (ACIS)}, \facility{Keck: Interferometer (LRIS)}, \facility{Swift (XRT, BAT)}, \facility{Effelsberg, \facility{ING:Herschel}}.

\bibliography{plane_blazar}

\end{document}